\documentstyle[a4,12pt]{article}
\begin{document}
\newcommand{\C}{C^{z}}
\newcommand{\Hb}{H^{z}_{\bar{\theta}}}
\newcommand{\D}{D_{\theta}}
\newcommand{\Db}{D_{\bar{\theta}}}
\newcommand{\zet}{\zeta_{\theta}}
\newcommand{\der}{\partial_{z}}
\newcommand{\R}{R_{z\theta}}
\newcommand{\covder}{\hat{\nabla}_{\zeta}}
\newcommand{\Del}{\Delta_{\zeta}}
\newcommand{\Delk}{\Delta_{\chi}}
\newcommand{\Delko}{\Delta_{\chi_{0}}}
\newcommand{\F}{\Phi_{\zeta}}
\newcommand{\Fk}{\Phi_{\chi}}
\newcommand{\Fko}{\Phi_{\chi_{0}}}
\newcommand{\zetp}{\zeta^{\prime}}
\newcommand{\demi}{\frac{1}{2}}
\newcommand{\dl}{d\lambda}
\newcommand{\dlb}{d{\bar{\lambda}}}
\newcommand{\M}{d^{2}\lambda}
\newcommand{\sig}{\hat{\Sigma}}
\newcommand{\dsig}{\partial\hat{{\cal D}}(\eta_{0})}
\newcommand{\fact}{\frac{1}{2i\pi}}
\newcommand{\facth}{\frac{1}{8i\pi}}
\newcommand{\kio}{\chi_{0}}
\newcommand{\Rko}{R_{\chi_{0}}}
\newcommand{\Tb}{\bar{T}_{\bar{\lambda}}}
\newcommand{\Pz}{P_{0k}}
\newcommand{\zt}{\tilde{z}}
\newcommand{\tetil}{\tilde{\theta}}
\newcommand{\Zh}{\hat{Z}}
\newcommand{\Teth}{\hat{\Theta}}
\newcommand{\zett}{\zeta_{T}}
\newcommand{\Pb}{\bar{P}}
\newcommand{\Dec}{\bigtriangledown}
\renewcommand{\baselinestretch}{1.2}
\def\scf{superconformal }
\def\scfty{superconformality }
\def\sd{superdiffeomorphism }
\def\sc{supersymmetric }
\def\sdf{superdifferential }
\def\sdfs{superdifferentials }
\def\sz{supersymmetrization }
\title{{\bf Induced Polyakov supergravity on Riemann surfaces of higher
genus}}
\author{{\bf Jean-Pierre Ader and Hamid Kachkachi} \\
{\em Laboratoire de Physique Th\'{e}orique, CNRS$^{\dag}$} \\
{\em Universit\'{e} de Bordeaux I, }\\
{\em 19 rue du Solarium, F-33175 Gradignan Cedex, France}}
\begin{titlepage}
\maketitle
\thispagestyle{empty}
\vspace{1 cm}
\begin{abstract}
 An effective action is obtained for the $N=1$, $2D-$induced supergravity on a
 compact super Riemann surface (without boundary) $\hat\Sigma$ of genus
 $g>1$, as the general solution of the corresponding superconformal Ward
 identity. This is accomplished by defining a new super integration theory on
 $\hat\Sigma$ which includes a new formulation of the super Stokes theorem
 and residue calculus in the superfield formalism. Another crucial
 ingredient is the notion of polydromic fields. The resulting action
 is shown to be well-defined and free of
 singularities on $\sig$. As a by-product, we point out a morphism between
 the diffeomorphism symmetry and holomorphic properties.
\end{abstract}
\vspace{4cm}
LPTB 93-10\\
hep-th/9310163\\
September 1993
Email addresses: {\bf Ader@FRCPN11.IN2P3.FR}, {\bf Kachkach@FRCPN11.IN2P3.FR}\\
$\dag$ $\overline{\mbox{Unit Associe au CNRS}}$, U.A.764\\ \\
\end{titlepage}
\section{Introduction}
In some recent publications \cite{AGN,G1} results concerning the factorization
of partition functions in superstring theories with non-vanishing central
charge were obtained. This supersymmetric generalization of well-established
facts on holomorphic factorization in string theory \cite{KLS} consists in
particular
in substracting from the effective action a local counterterm that allows one
to go from the super Weyl to the superdiffeomorphism anomaly defined
on a compact super Riemann surface (SRS) without boundary. The characteristic
feature of the latter anomaly is that it appears as the sum of two
contributions
corresponding respectively to the holomorphic and antiholomorphic sectors
of the theory. Furthermore it has been shown \cite{G1} that, for a generic
value of the central charge, the holomorphic factorization of partition
functions remains true for free superconformal fields when these functions
are considered as functionals of the Beltrami coefficients and their
fermionic partners.

 Until now a generalization of these studies to a generic
 SRS has been worked out only for the supertorus $(g=1)$\cite{AK}, providing
 thereby an explicit expression for the Polyakov action thereon.
 Here we extend these results to a compact SRS (without boundary) of arbitrary
 genus. The demonstration and formulation obtained are similar in spirit
 to those of the bosonic case \cite{Z1,Z2} but encounter considerable
 complications inherent to the superspace formulation. Some difficulties
 have been overcome through the systematic use of covariant derivatives and
 ``tensor'' notations. These technical tools are not mandatory in the bosonic
 case but become indispensable in the present more complicated situation.
 The other difficulty comes from the use of objects which can be singular
 on the SRS and require a careful study of their analytic properties.
 Obviously it is always possible, in principle, to consider their component
 expansion expressing these superfields in terms of usual fields which possess
 well-known analytic properties on the underlying Riemann surface. However
 this
 path can be technically very complicated and in fact renders the superfield
 approach useless and rather artificial. On the contrary, as we will show,
 it is possible to derive the general Polyakov action on a SRS of any genus
 entirely in terms of superfields throughout the calculations.

 As a by-product, we have obtained a super analog of Stokes theorem
 (subsect.2.2) allowing us to perform any integral (when only integrable
 singularities are present)
 over a compact SRS by using super Cauchy theorems, without having recourse to
 component expansion. Indeed, this was possible by defining a new super
 integration procedure, which is not only suitable for our purpose but is
 interesting in its own right as well, and could be used for further
 generalizations.
 Furthermore, we have noticed a canonical morphism between the
 superdiffeomorphism transformations and holomorphy properties of the
 superfields on a SRS (sect.3).

 Let us first review the properties of the $N=1$ compact SRS  $\sig$ to
 establish the framework in which we are working. This surface is locally
 described by a pair of complex coordinates $(z,\theta)$  with $\theta$
 anticommuting \cite{FD},
 and glued up from local charts by superconformal transition
 functions, i.e. two local coordinate charts are related
 by transformations that satisfy the superconformal condition
 \footnote{Obviously it is understood that the complex conjugate (c.c)
 conditions are also taken into account.} $\D\zt=\tetil\D\tetil$, where
 $\D = \partial_{\theta}+\theta\partial_{z}$ is the superderivative. Here
 $\sig$ is of De Witt type, i.e. a SRS to which is associated a
 corresponding compact Riemann surface $\Sigma$ of genus $g$, called its body,
 with a particular spin structure. Such a SRS is adequate for a picture of a
 moving superstring in spacetime\cite{RC}.

 In addition to the reference supercomplex structure $\{(z,\theta)\}$, a
 SRS can be provided with the so-called projective structure. This is a
 collection of local homeomorphisms $(\Zh_{\alpha},\Teth_{\alpha})$ of
 $\sig$ into $I\! \! \!$C$^{1\vert1}$, obeying the gluing laws
 \footnote { The matrix $ \left( \begin{array}{c c} a&b \\ c&d \\ \end{array}
 \right)$ belongs to $SL(2,I\! \! \!$C) whereas $\alpha$ and $\beta$ are odd
  Grassmann numbers.}
 on overlapping domains \cite{RC}

\begin{eqnarray}\label{1}
\tilde{\Zh}&=&\frac{a\Zh+b}{c\Zh+d}+\Teth\frac{\alpha\Zh+\beta}{(c\Zh+d)^{2}}
\nonumber \\
\tilde{\Teth}&=&\frac{\alpha\Zh+\beta}{c\Zh+d}+\Teth\frac{1}{c\Zh+d}
(1+\demi\alpha\beta)
\end{eqnarray}
This structure is related to the reference structure $\{(z,\theta)\}$ by a
quasisuperconformal transformation on $\sig$ which is parametrized by a pair
of fields,
the super Beltrami coefficient $\Hb$ ( which contains the ordinary Beltrami
coefficient ) and the super Schwarzian derivative $S(\Zh,\Teth;z,\theta)$
( and c.c ) \cite{DG}. In fact the number of independent Beltrami
coefficients is greater than one; the parametrization above corresponds to
a particular choice which  has been proved to be always possible \cite{Tak}
and is adopted now. This choice corresponds to the special case
$H^{z}_{\theta}=0$  and is equivalent to the condition $\D\Zh=\Teth\D\Teth$.
Henceforth, unless otherwise stated, holomorphy will be understood with
respect to the projective structure $\{(\hat{Z},\hat{\Theta})\}$, this is
sometimes referred to as the $H-$structure, while the reference structure
will be referred to as the $0-$structure. Objects that are holomorphic with
respect to the latter structure are indexed by a subscript 0.

\section{The Polyakov action on a SRS}

The superspace generalization of Polyakov's chiral gauge action proposed by
Grundberg and Nakayama \cite{GN} for the planar topology is

 \begin{equation} \label{a}
 \Gamma[\Hb]=\int_{\bf SC}\M\der\zet\Hb
 \end{equation}

 where $\zet$ is the coefficient of a super affine connection
 built out of the solution of the super Beltrami equation $(\Zh,\Teth)$
 as follows
 \begin{equation}\label{B}
 \zet=-\D\ln\D\Teth.
 \end{equation}

 This allows us to build a superprojective connection \cite{Gie}
\begin{equation} \label{c}
\R= -\der\zet-\zet\D\zet
\end{equation}

 which under a coordinate transformation behaves like a super Schwarzian
 derivative.

 The measure in eq.(\ref{a}) reads
$$\M=\frac{\dl\wedge\dlb}{2i}$$
where $\dl=(dz\vert d\theta)$ is the generator of the supercanonical line
bundle over $\sig$ whose body is a spin bundle over the underlying Riemann
surface $\Sigma$ \cite{Nel}. The generator $\dlb$ is defined in a similar way.

The covariant superderivative $\covder$ associated to the superaffine
connection $\zet$ is defined by its action on superfields or operators of
conformal weights $p$ corresponding to the $(z,\theta)$ sector
( for instance, $p(\Hb)=-1$, $p(\D)=\demi$) by\cite{Gie}

\begin{equation} \label{d}
\covder=\D+2p\zet
\end{equation}

and consequently the covariantization of $\der$ is given by

\begin{equation} \label{e}
\Del=\der+2p\D\zet+\zet\D.
\end{equation}

When acting on itself or on another super affine connection $\zetp$
 the covariant derivative defines the corresponding ``field strength''
\begin{equation} \label{f}
\F\zetp=\D\zetp+p(\zetp)\zetp.
\end{equation}

Remembering that $\zet$ and $\zetp_{\theta}$ are superfields of type
$(\demi,0)$ and by applying once more the covariant superderivative,
we get ($\zetp = \zetp_{\theta}$ )

\begin{equation} \label{g}
\Del\zetp=\der\zetp+\demi\zet\D\zetp+\demi\zetp\D\zet
\end{equation}

 which is the covariantization of $\der\zetp$; here the derivative of the
 superaffine connection has been changed ( up to a sign ) into a
 superprojective connection.

 So far, a proper generalization of eq.(\ref{a})  to higher genus SRS has not
 been found except for the case of the supertorus \cite{AK}. The integrand
 of the globally defined super Polyakov action for genus $g=1$ is given by

\begin{equation}\label{h}
A_{ST}=4[(\R+\Delta_{\zet}\zet)\Hb+\demi(\zet-\zeta_{0})
\Delta_{\zett}\Hb]
\end{equation}

where $\zet$ is the coefficient of a superaffine connection on the supertorus,
 given by the same equation as (\ref{B}) and $\zeta_{0}\equiv \zeta_{0\theta}$
the holomorphic superaffine connection in the reference structure.

This expression solves the Ward identity
\begin{equation} \label{j}
s(\Gamma[\Hb,\R]+ c.c)=\int_{ST} \M[{\cal A}(\C,\Hb,\R)+ c.c]
\end{equation}

 where s is the BRST operator \footnote{In superspace the s-operator is
 assumed to act as an antiderivation from the right; the BRST algebra is
 graded by the ghost number, but does not feel the Grassman parity.} and
 ${\cal A}(\C,\Hb,\R)$ is the chirally split form of the globally defined
 (non integrated ) superdiffeomorphism anomaly \cite{DGP}

\begin{eqnarray} \label{k}
{\cal A}(\C,\Hb,\R)&=&\C\der^{2}\D\Hb+\Hb\der^{2}\D\C
+3\R(\C\der\Hb-\Hb\der\C)   \nonumber \\
		 &+&\D\R(\C\D\Hb+\Hb\D\C).
\end{eqnarray}

 $\C$ is the \sd   ghost field and $\R$ is a holomorphic superprojective
 connection which renders the expression above well-defined.

Now our starting point for the further generalization of (\ref{h}) to a SRS
of genus $g>1$ is the following functional which is inspired from the one
found in the bosonic case in ref.\cite{Z1}.
\begin{equation} \label{a1}
\Gamma_{1}[\R,\Hb]=\frac{1}{4\pi}\int_{\sig}\M\hat{A}_{1}
\end{equation}
 where
\begin{eqnarray} \label{a2}
\hat{A}_{1}&=&4\R\Hb+2\der\D\Hb+2\kio\D\chi\Hb+2\chi\D\kio\Hb  \nonumber \\
	   &-&\D\kio\D\Hb-\D\chi\D\Hb-\kio\der\Hb-\chi\der\Hb
\end{eqnarray}

  This expression can also be written in a compact form as

\begin{equation} \label{a3}
\hat{A}_{1}=4(\R-\Rko)\Hb+\D(\Delk+\Delko)\Hb
\end{equation}
 which thus becomes obviously globally defined.

 Here, the coefficients $\chi$  and $\chi_{0}$ are the superaffine connections
 related to the polydromic $\demi$-superdifferentials $\Psi$, $\Psi_{0}$ by

\begin{equation}\label{a4}
\chi=-\D\ln\Psi  \hspace{1cm} ,  \hspace{1cm}  \chi_{0}=-\D\ln\Psi_{0}.
\end{equation}

 The coefficients $\Psi$, $\Psi_{0}$ are superholomorphic differentials in
 the H-structure and 0-structure respectively. They are free of zeros and
 consequently are multivalued objects on $\sig$. $\Rko$ is a particular
 holomorphic superprojective connection in the H-structure related to
 $\chi$, $\chi_{0}$ by

 \begin{equation} \label{a5}
\Rko=\demi(\Delk\chi_{0}+\Delko\chi)
\end{equation}
 Let us now discuss these

 \subsection{Polydromic fields}

 According to the super Riemann-Roch Theorem \cite{Nel,RSV}
 \footnote{See in particular \cite{RSV} for a proof in both cases of a split
 and non-split SRS.}, there are $g$ holomorphic single-valued
 $\demi-$superdifferentials which have each
 globally $(g-1)$ (counting multiplicity) zeros on a compact SRS of genus $g$.
  We shall consider in the following two kinds of these differentials;
  the holomorphic
 ones w.r.t. the $H-$structure will be denoted by $\eta$ and
 the holomorphic objects in the $0-$structure by $\eta_{0}$.
 In contrast, the multivalued \sdfs  $\Psi, \Psi_{0}$ considered in
 eq.(\ref{a4}) are free of zeros. They can be written, in general, as
 \cite{K2}

 \begin{equation} \label{a6}
 \Psi=(\D\Teth) \xi (\hat Z,\hat\Theta)  \hspace{1cm} ,  \hspace{1cm}
	\Psi_{0}=(\D\Teth_{0}) \xi_{0}' (z,\theta)
\end{equation}

 where $\Teth_{0}$ is the Grassmann coordinate of another reference structure
related to the $0-$structure by a superconformal transformation.
 The expressions given above  indeed transform as $\demi-$\sdfs$ $ since
 $\D\Teth$ does. The functions $\xi(\hat Z,\hat\Theta)  ,
 \;\xi_{0}'(z,\theta)$ are
 multivalued and no-where vanishing \cite{K2}.

However, to deal with these differentials, there are two equivalent methods.
The first one considers these objects as defined on a universal covering of
the SRS and invariant under the corresponding covering group.
 Then one chooses a fundamental domain where these differentials
 ( in fact their coefficients ) become single-valued. This insures that
 the whole expression they appear in is single-valued. However a second
 approach consists in choosing a branch for
 each of these differentials by working on a dissection of the SRS into its
 polygon. Mimicking the bosonic case \cite{Z1} we cut the SRS $\sig$  into
 its polygon whose reduced domain is the polygon of the underlying Riemann
 surface. This dissection has $4g$ pairwise opposite sides and will be
 denoted by $\hat{\cal D}$. This is assumed to contain all zeros of
 $\eta_{0}$ while excluding those of $\eta$. In the bosonic case, this was
 shown to be always possible \cite{Z1}, and even when zeros of the
 (corresponding) bosonic differentials $\omega$ and $\omega_{0}$ overlap, the
 total Polyakov action is still continuous
 \footnote{We are indebted to R. Zucchini for valuable discussions on this
 subject and in particular on polydromic differentials.}\cite{Z2}. Using the
 trivial topology of De Witt, one can repeat the same demonstration on $\sig$.
 Next we cut infinitesimal
 superloops ${\cal C}_{k}$ around the zeros $P_{0k}$ of $\eta_{0}$ with
 opposite orientation in regard to that of the boundary $\partial\hat{\cal D}$
 of
 $\hat{\cal D}$. These superloops are composed of their bodies
 ( ordinary circles in $\Sigma$ ) which define the corresponding orientation
 and some ``Grassmann circles'' ( defined below ) over them.

 The s-variation of a superaffine connection \cite{AK} is simple when
 expressed in terms of covariant derivatives of the ghost field $\C$

\begin{equation} \label{b1}
s\chi=-\demi\D\Delk \C
\end{equation}

 whereas other useful transformation laws are

\begin{equation} \label{c1}
s(\Delk\Hb)=\Db\Delk\C \hspace{1cm} ; \hspace{1cm}  s(\Delta_{\chi}\chi)=
 \demi\Delk\D\Delk\C.
\end{equation}

 Now since the superaffine connection $\chi_{0}$ is s-invariant, using the
 above laws and the BRST transformation of $\Hb$ given in Ref.\cite{DG} we
 obtain, after lengthy but straightforward calculations

\begin{equation} \label{d1}
s\Gamma_{1}[\R,\Hb]=\frac{1}{4\pi}\int\M{\cal A}(\C,\Hb,\R)+K_{1}+K_{2}+K_{3}
\end{equation}

 where

\begin{eqnarray} \label{e1}
 K_{1}&=& \fact\oint_{\dsig}\dl [(\R-\Rko)\C]  \nonumber\\
 K_{2}&=& -\facth\oint_{\dsig}\dlb (\Delk s\Hb+\Delko\Db\C+
 2\C\D\Db\chi+\D\C\Db\chi) \nonumber\\
 K_{3}&=& -\facth\oint_{\dsig}\dlb [\demi(\chi_{0}-\chi)\D(\D\Hb\D\C)
 -\chi_{0}\chi(\D\Hb\D\C)   \nonumber\\
      & &   -(\chi_{0}\D\chi+\der\chi-\chi_{0}\chi\D+\R)(\C\D\Hb+\Hb\D\C)]
 \end{eqnarray}

   and ${\cal A}(\C,\Hb,\R)$ is the globally defined non integrated anomaly
 already given by the same expression as in eq.(\ref{k}) but now defined on
 $\sig$.

First we wish to stress that the terms in eq.(\ref{e1}) are separately
globally
 defined. This is obvious for $K_{1}$ since its integrand is the product of
 the difference of two superprojective connections (i.e. a superquadratic
 differential) and $\C$ which transforms homogeneously. Concerning $K_{2}$ and
 $K_{3}$ this property has to be checked by hand. While $K_{2}$ contains the
bosonic limit ( see ref.\cite{Z1}, eq.(2.15)), the fourth
term $K_{3}$ gathers the complications and novelties emerging from the
 supersymmetric formulation. Note that all these terms are in fact untractable
 because they all involve multivalued fields. Nevertheless, we will show
 later that these terms can be eliminated by adding other contributions to the
 Polyakov action, thus leaving the anomaly only.

To explain the line integrals in eqs.(\ref{e1}) we now present our integration
procedure over a (compact) SRS with respect to which our solution to the
superconformal Ward identity (\ref{j}) will be defined. Most importantly, we
give the definition of the boundary of a superdomain $\hat{\cal D}$ in $\sig$,
and the analog of Stokes theorem to relate integration over superdomains on
$\sig$ and integration on their boundaries. \\

\subsection{Integration on $\sig$}

In our developments, the expression that we integrate over $\hat\Sigma$
is a $(\frac{1}{2},\frac{1}{2})-$superdifferential
which is, in general, meromorphic . More precisely, this expression
happens to be a function of singular objects like $\partial\log\eta_{0}$
( or $\D\log\eta_{0}$) inside a domain $\hat{\cal D}$ containing all zeros of
$\eta_{0}$. To perform explicitly the corresponding integral and in
particular the residue calculus, we first need an analog of Stokes theorem
and a consistent procedure of integration on the boundary
$\partial\hat{{\cal D}}$ of $\hat{{\cal D}}$. In fact, integration on
$\partial\hat{{\cal D}}$ reduces to the sum of integrals over small
``circles'' ${\cal C}_{k}$ surrounding the zeros
$P_{0k}=(z_{0k},\theta_{0k})$ of
$\eta_{0}$; the orientation of these circles being opposite to that of
$\partial\hat{{\cal D}}$.
Since the remaining integration path is a sequence of pairs of
geometrically coinciding but oppositely oriented arcs, this yields pairs of
mutually cancelling contributions when the integrand is single-valued.
Summing up, the only remnant is the contribution coming from the
${\cal C}_{k}$'s. As explained above, these circles are composed of ordinary
circles in $\Sigma$ and some ``Grassmann circles'' ${\cal C}_{\theta}$
around the singular point
$\theta = \theta_{0},\; \bar\theta=\bar\theta_{0}$. ${\cal C}_{\theta}$
is defined in such a way that the identity
\begin{equation}\label{new}
 \int_{{\cal C}_{\theta}}d\theta f(z,\theta,\bar z,\bar\theta)=
 (\partial_{\theta}f)(z,\theta_{0},\bar z,\bar\theta_{0})
\end{equation}
holds for every (locally) smooth function $f$ on $\hat\Sigma$.\\
Thus any integration over Grassmann numbers is performed over
${\cal C}_{\theta}$ instead of the whole space of Grassmann variables as is
done in the standard Berezin integration.
The usual Berezin rules obviously hold (in particular) on ${\cal C}_{\theta}$
\begin{equation}\label{Br1}
\int_{{\cal C}_{\theta}}d\theta = 0,\hspace{1cm}
\int_{{\cal C}_{\theta}}d\theta(\theta-\theta_{0}) = 1,\hspace{1cm}
\int_{{\cal C}_{\theta}}d\theta(\bar\theta-\bar\theta_{0}) = 0.
\end{equation}

However, our integration procedure (\ref{new}) marks a crucial departure from
that of Berezin by the result
\begin{equation}\label{Br3}
\int_{{\cal C}_{\theta}}d\theta(\theta-\theta_{0})(\bar\theta-\bar\theta_{0})
 = 0
\end{equation}
Indeed, in our point of view $\bar\theta$ is treated somehow as the complex
conjugate of $\theta$ and hence $\theta$ (or $\bar\theta$) can not be taken
out of the (line) integral over $\bar\theta$ (or $\theta$), since these
variables are in fact linked on ${\cal C}_{\theta}$. This super integration
formalism reproduces the results obtained by the component expansion while
avoiding the well-known difficulties of this procedure, especially when the
superfields are singular. Obviously, the rules (\ref{new})-(\ref{Br3}) are
the same if ${\cal C}_{\theta}$ is a circle around the point
$\theta_{0}=0, \bar\theta_{0}=0$.

The rule (\ref{Br3}) (in addition to others ) is, as we will show below,
crucial to establish the super Stokes theorem, which thereby reads
\footnote{See \cite{BL,VZ} for other formulations of this theorem.}

\begin{equation}\label{St}
\int_{\hat{\cal D}}\hat{d}\Phi = \oint_{\partial\hat{{\cal D}}}\Phi
\end{equation}
where the coboundary operator $\hat{d}$ is defined by its action on a
$(p/2,q/2)-$superdifferential $\Phi$ as follows
\footnote{$\hat{d}$ will be denoted $d_{+}$ or $d_{-}$ when $(p+q)$ is even
or odd respectively.}
\begin{equation}\label{zz}
\hat{d}\Phi=(\dl\D + (-1)^{(p+q)}\dlb\Db)\Phi.
\end{equation}

It is  straightforward to see that the operator $\hat{d}$ is nilpotent
as it must be, i.e.\footnote{Here $\Phi$ is explicitly
written as $\Phi =\Phi_{\theta}(\dl)^{p}(\dlb)^{q}$ and the operators $\D$
and $\Db$ act directly on the coefficient $\Phi_{\theta}$.
We find $d_{+}d_{-}=-d_{-}d_{+}=0$.}$\hat{d}^{2} = 0$.

Now let us check that the theorem (\ref{St}) is indeed based on the
integration rules (\ref{new})-(\ref{Br3}). First we recall that the
expression that can be integrated over a SRS is a
$(\demi,\demi)-$ superdifferential, and
hence in (\ref{St}) $\Phi$ is a linear combination of $(\demi,0)-$ and
$(0,\demi)-$superdifferentials. So, let us simply consider a
$\demi-$superdifferential $\Phi = \Phi_{\theta} \dl,$ then expand the even
Grassmann coefficient
$\Phi_{\theta}$ in its $\theta$, $\bar\theta$ components around the point
$\theta=\theta_{0}, \;\bar\theta=\bar\theta_{0}$, that is
\begin{equation}\label{Exp}
\Phi_{\theta}(z,\theta,\bar{z},\bar\theta) = \phi_{0} +
(\theta - \theta_{0})\phi_{1} + (\bar\theta - \bar\theta_{0})\phi_{2} +
(\theta - \theta_{0}) (\bar\theta - \bar\theta_{0}) \phi_{3}.
\end{equation}
This yields
\begin{eqnarray*}
\hat{d}\Phi&=&(d_{-}\Phi_{\theta})\wedge\dl = (\Db\Phi_{\theta})\dlb\wedge\dl
\\&=& [\phi_{2} - (\theta - \theta_{0})\phi_{3} +
(\bar\theta - \bar\theta_{0})\bar{\partial}\phi_{0} -
(\theta - \theta_{0}) (\bar\theta - \bar\theta_{0}) \bar{\partial}\phi_{1}]
\dlb\wedge\dl.
\end{eqnarray*}
Now due to the usual Berezin integration rules (\ref{Br1}), we get
\begin{eqnarray*}
\int_{\hat{\cal D}}\hat{d}\Phi&=&-\int_{\hat{\cal D}}\dlb\wedge\dl
(\theta - \theta_{0}) (\bar\theta - \bar\theta_{0}) \bar{\partial}\phi_{1}
= -\int_{{\cal D}}d\bar{z}\wedge dz \bar{\partial}\phi_{1}\\
&=& \oint_{\partial{\cal D}}\phi_{1}dz
\end{eqnarray*}
where in the last step we used the usual Stokes theorem on ${\cal D}$, the
underlying domain of $\hat{\cal D}$.\\
The right hand side of (\ref{St}) is readily computed using now the rules
(\ref{new})-(\ref{Br3}),
$$\oint_{\partial\hat{{\cal D}}}\Phi =  \oint_{\partial{\cal D}}\phi_{1}dz$$
thus showing that the left and right hand sides of (\ref{St}) indeed
coincide.\\
Here we wish to emphasize that this theorem is actually more subtle than it
appears to be, since it allows us to compute integrals in the superfield
formalism and thus spares us the generally cumbersome procedure of expanding
superfields in their components, especially when these are singular, since
there is no telling, in general, which components exhibit the corresponding
singularities.

Now using these integration rules for Grassmann variables, the integral of a
meromorphic $(\frac{1}{2},0)-$ or $(0,\frac{1}{2})-$superdifferential $\Phi$
over the circles ${\cal C}_{k}$ is performed by using the
(generic) local behaviour around $P_{0k}$\cite{K2}
\footnote{N is the number of terms in $\Phi$.}
\begin{equation}\label{Lb}
\Phi\sim \sum_{j}^{N}\frac{f_{j}}{(z-z_{0k}-\theta\theta_{0k})^{j}}
\end{equation}
where the coefficient functions $f_{j}$ are superholomorphic around $P_{0k}$;
i.e. they do not depend on $(\bar z,\bar\theta)$ inside an open neighbourhood
of $P_{0k}$ contained in ${\cal C}_{k}$. Then we get the final result
with the help of the two super Cauchy theorems\cite{FD}
\begin{eqnarray}\label{y1}
\fact\oint_{{\cal C}_{k}}\dl f(z,\theta)(z-z_{k}-\theta\theta_{k})^{-n-1}
 &=& \frac{1}{n!}\partial_{z_{k}}^{n}D_{\theta_{k}}f(z_{k},\theta_{k})
 \nonumber\\
\fact\oint_{{\cal C}_{k}}\dl f(z,\theta)(\theta-\theta_{k})(z-z_{k}-\theta
\theta_{k})^{-n-1} &=& \frac{1}{n!}\partial_{z_{k}}^{n} f(z_{k},\theta_{k})
\end{eqnarray}

\subsection{Other contributions to the Polyakov action and residue calculus}

 From the experience gained in the construction of the Polyakov action on
 an ordinary Riemann surface \cite{Z1}, we know that at least two other terms
 are necessary to describe the super Polyakov action for any genus g, namely
 one which has to be a $(0,\demi)$-\sdf  and another which makes explicit some
 residues on the dissected SRS so as to cancel out the terms
 $K_{1}, K_{2}, K_{3}$ in eqs.(\ref{d1}).

 Therefore we introduce the following contribution to the Polyakov
 action

\begin{equation} \label{f1}
\Gamma_{2}[\Hb]=\fact\oint_{\dsig}\ln(\Psi/\eta_{0})\hat{d}
\ln(\eta/\Psi_{0})\equiv\fact\oint_{\dsig}\hat{A}_{2}
\end{equation}
Then under the following BRST transformation laws
\begin{equation}\label{tl}
s\ln\Psi=\demi\Delk\C, \; s\Psi_{0}=0,\;
s\ln\eta=\demi\Delta_{\zeta}\C, \; s\eta_{0}=0
\end{equation}
the response of the functional $\hat{A}_{2}$ reads

\begin{equation}\label{h1}
 s\hat{A}_{2}=\hat{d}[s\ln\eta\ln(\Psi/\eta_{0})]
		+ \dl T_{\lambda}+\dlb\bar{T}_{\bar{\lambda}}
\end{equation}

 where
\begin{eqnarray} \label{i1}
T_{\lambda}&=&\demi\D\phi+(\R-R_{\zeta_{0}})\C-(\R-\Rko)\C \hspace{1cm}(a)
\nonumber\\
\bar{T}_{\bar{\lambda}}&=&s\ln\Psi\Db\ln(\eta/\Psi_{0})
			-s\ln\eta\Db\ln(\Psi/\eta_{0})\hspace{3.3cm}(b)
\end{eqnarray}
and
\begin{equation} \label{j1}
\phi=\D[\C(\chi_{0}+\chi-\zeta_{0}-\zeta)]+(\zeta\zeta_{0} - \chi\chi_{0}
	-2\zeta\chi)\C
\end{equation}

 is globally defined (i.e $\tilde{\phi}\equiv\phi$ ). $\zeta_{0}$ and $\zeta$
 are the superaffine connections built out of $\eta_0$ and $\eta$ respectively
\begin{eqnarray*}
\zeta_0 =-\D\ln\eta_0  \hspace{1cm} ,  \hspace{1cm}  \zeta=-\D\ln\eta.
\end{eqnarray*}

Note that the projective connection $\R$ in (\ref{i1}a) is needed to ensure
the appropriate gluing of the integrand and may be replaced by any other
superprojective connection which is holomorphic and single-valued on the
integration domain.

 $T_{\lambda}$ is globally defined for the same reason as for the term
 $K_{1}$ in (\ref{e1}), whereas $\bar{T}_{\bar{\lambda}}$ is globally
 defined because it only involves quotients of $\demi-$differentials, i.e.
 invariant functions\footnote{The $s-$variation does not change the rules
 of gluing, since it leaves the variables $z,\;\theta$ invariant.}.\\
 Using the holomorphy equations satisfied by $\Psi$, $\Psi_{0}$, $\eta$,
 $\eta_{0}$
\begin{equation} \label{k1}
\Db\ln\Psi=\demi\Delk\Hb  \hspace{.5cm} , \hspace{.5cm} \Db\Psi_{0}=0
  \hspace{.5cm} , \hspace{.5cm} \Db\ln\eta=\demi\Del\Hb  \hspace{.5cm}
  , \hspace{.5cm}  \Db\eta_{0}=0
\end{equation}

 we obtain

\begin{equation} \label{l1}
\Tb=\frac{1}{4} (\Delk\C\Del\Hb-\Del\C\Delk\Hb).
\end{equation}

 We note that since the expression $\phi$ (which is globally defined ) appears
 in $T_{\lambda}$ as the holomorphic part of a total derivative, we can get
 rid of the first term in eq.(\ref{i1}a)  by completing it in such a way that
 it becomes
 a total derivative. This is indeed achieved by adding and substracting the
 term $\demi\dlb\Db\phi$ in $s\hat{A}_{2}$. Thus we obtain
 \begin{equation}
 s\hat{A}_{2} = \hat{d}[\frac{s\eta}{\eta}\ln(\Psi/\eta_{0}) +
		\demi\phi] + \dl T^{'}_{\lambda}+\dlb \bar{T}^{'}_
{\bar\lambda}
\end{equation}

where
$$T^{'}_{\lambda}= (\R-R_{\zeta_{0}})\C-(\R-\Rko)\C$$
and
\begin{equation}\label{m1}
\Tb'=\Tb-\demi\Db\phi.
\end{equation}

 Now the whole expression on which
 $\hat{d}$ acts is globally defined and single-valued and thus vanishes when
 we integrate along infinitesimal circles surrounding the zeros of
$\eta_{0}$.\\
 Explicitly $\Tb'$ reads
\begin{eqnarray} \label{n1}
\Tb'&=&\frac{1}{4}\{(\Delk s\Hb+\Delko\Db\C+2\C\D\Db\chi+\D\C\Db\chi)
\nonumber  \\
&-&(\Del s\Hb+\Delta_{\zeta_{0}}\Db\C+2\C\D\Db\zeta+\D\C\Db\zeta)
\nonumber  \\
&+&\D(\chi-\zeta)\D\Hb\D\C+\demi(\zeta-\chi)\D(\D\Hb\D\C) \nonumber  \\
&+&(\zeta-\chi)(\C\der\D\Hb+\Hb\der\D\C)+\D\C(\Db\chi-\Db\zeta)  \nonumber  \\
&+&(\chi+\kio-\zeta-\zeta_{0})\D\Db\C-2\Db[\C(\zeta\zeta_{0}-\chi\kio-
2\zeta\chi)]  \nonumber\\
&+&2(\chi\D\zeta-\zeta\D\chi)(\C\D\Hb+\Hb\D\C)+2\zeta\chi\D\Hb\D\C)\}.
\end{eqnarray}

At this point we note that the second term in $T^{'}_{\lambda}$ cancels out
$K_{1}$ in eq.(\ref{d1}), and the first line in $\Tb'$ cancels out $K_{2}$.
However, to get rid of $K_{3}$ and the other terms in $\Tb'$, we still need
to add the following functional

\begin{equation} \label{p1}
\Gamma_{3}[\Hb]=\frac{1}{4i\pi}\oint_{\dsig}\dlb [(\zeta\zeta_{0}
		-\chi\chi_{0}-2\zeta\chi)\Hb-\demi(\chi+\chi_{0}-\zeta
		 -\zeta_{0})\D\Hb].
\end{equation}

 It is easy to check that the integrand of this expression is globally
 defined. When expanded in components \cite{AK} this
 contribution disappears in the bosonic limit.

  Therefore, the combination of all these contributions into
  $\Gamma=\Gamma_{1}+\Gamma_{2}+\Gamma_{3}$ yields after several ( trivial )
  cancellations
\begin{eqnarray} \label{q1}
s\Gamma=\int{\cal A}(\C,\Hb,\R)&+&\frac{1}{8i\pi}\oint_{\dsig}
\{\dlb[\R(\C\D\Hb+\Hb\D\C)+2(2\R\C+\der\zeta\C)  \nonumber \\
&+&\demi(\zeta-\zeta_{0})\D(\D\Hb\D\C)+\zeta_{0}\zeta\D\Hb\D\C  \nonumber \\
&+&(\zeta_{0}\D\zeta-\zeta_{0}\zeta\D+\der\zeta)(\C\D\Hb+\Hb\D\C)]\nonumber\\
&-&(\Del s\Hb+\Delta_{\zeta_{0}}\Db\C+2\C\D\Db\zeta+\D\C\Db\zeta)\}
\nonumber\\
&+&\frac{1}{2i\pi}\oint_{\dsig}\dl[(\R-R_{\zeta_{0}})\C].
\end{eqnarray}

 A careful inspection shows that, since terms which are not singular at
 the zeros $\Pz$ of $\eta_{0}$ do not contribute, expressions which are
 likely to give non-zero results are those containing $\eta_{0}$. These
 reduce to
\begin{equation} \label{r1}
 J_{1}=\fact\oint_{\dsig}\dl[(\R-R_{\zeta_{0}})\C]
\end{equation}

\begin{equation}\label{r1b}
 J_{2}=-\frac{1}{8i\pi}\oint_{\dsig}\dlb [\Delta_{\zeta_{0}}\Db\C+
  \demi\zeta_{0}\covder(\D\Hb\D\C)-\zeta_{0}\D(\zeta(\C\D\Hb+\Hb\D\C))]
\end{equation}

 The integrand of $J_{1}$, being in fact $\demi[(2\R+\der\zeta)\C+
 (\der\zeta_{0}+\zeta_{0}\D\zeta+\zeta\D\zeta_{0})\C]$, reduces to
\begin{equation} \label{s1}
 J_{1}=\frac{1}{4i\pi}\oint_{\dsig}\dl (\der\zeta_{0}+\zeta_{0}\D
 \zeta+\zeta\D\zeta_{0})\C
\end{equation}
 since the other terms are non-singular in the integration domain

  Now $\eta_{0}$ can be locally written in a neighborhood ${\cal O}_{k}$
  of $\Pz$ as \cite{K2}

\begin{equation}\label{t1}
\eta_{0}(P_{k})=\beta(P_{k})(z_{k}-z_{0k}-\theta_{k}\theta_{0k})
^{\demi\alpha_{0k}}.
\end{equation}

 Here ($z_{k},\theta_{k}$) are the coordinates of the point $P_{k}$
 belonging to ${\cal O}_{k}$ and ($z_{0k},\theta_{0k}$) those of $P_{0k}$;
 $\beta$ is an even
 superholomorphic function nowhere vanishing on $\sig$. Therefore
 the local behaviour of $\ln\eta_{0}$ is

\begin{equation}\label{u1}
\ln\eta_{0}=\demi\alpha_{0k}\ln(z_{10})+\ln\beta ,
\end{equation}

 where the supercoordinate displacements $z_{10}, \theta_{10}$ are
 defined by

\begin{equation}\label{v1}
 z_{10}=z_{k}-z_{0k}-\theta_{k}\theta_{0k}    \hspace{1cm}  ,
\hspace{1cm}   \theta_{10}=\theta_{k}-\theta_{0k}
\end{equation}

This implies that locally around $P_{0k}$ we have\footnote{In the
neighborhood of any point $P$ we can pick a particular parametrization,
defined by the normalization $z_{0k}=\theta_{0k}=0$ \cite{K2}}

\begin{equation}\label{w1}
\zeta_{0}\simeq-\demi\alpha_{0k}\frac{\theta_{10}}{z_{10}}-\D\ln\beta.
\end{equation}

Note that since $\beta$ is nowhere vanishing, the term $\D\ln\beta$ here
is regular and single-valued and therefore it disappears upon integration
over the infinitesimal circles ${\cal C}_{k}$.
 Putting this behaviour back into $J_{1}$ in eq.(\ref{s1}), we get
 \footnote{Recall that a global (-) sign appears because the circles
 ${\cal C}_{k}$ have opposite orientation with respect to
 $\partial\hat{\cal D}(\eta_{0})$.}

\begin{equation}\label{x1}
 J_{1}=-\sum_k\frac{\alpha_{0k}}{4}\fact\oint_{{\cal C}_{k}}\dl(
\frac{\theta_{10}}{z_{10}^{2}}\C-\frac{\theta_{10}}{z_{10}}\C\D\zeta
-\frac{1}{z_{10}}\zeta\C)
\end{equation}
 and then using the super Cauchy 's theorems (\ref{y1}) we finally obtain
 \footnote{The use of these theorems here is justified by the fact
 that $\C$ and $\zeta$ are superholomorphic inside the ${\cal C}_{k}$'s, i.e.
 they don't depend on $(\bar z,\bar\theta)$ therein.}
\begin{equation}\label{z1}
J_{1}=-\sum_k\frac{\alpha_{0k}}{4}\Del\C(P_{0k}).
\end{equation}
This term can then be readily cancelled out in (\ref{q1}) by adding the last
contribution to the Polyakov action

\begin{equation}\label{ab}
\Gamma_{4}[\Hb]=\demi\sum_k\alpha_{0k}\ln(\eta/\Psi_{0})(P_{0k}),
\end{equation}
since $$s\Gamma_{4}[\Hb]= \sum_k\frac{\alpha_{0k}}{4}\Del\C(P_{0k})$$ due to
(\ref{tl}).

Let us now turn to the remaining integral $J_{2}$. The first term
$$I = -\frac{1}{8\pi i}\oint_{\dsig}\dlb\Delta_{\zeta_{0}}\Db\C$$ is computed
using the local behaviour (\ref{t1}) of $\eta_{0}$ and discarding the
non-singular term $\partial\Db\C$

$$I=\frac{1}{8i\pi}\sum_k\alpha_{0k}\oint_{{\cal C}_{k}}\dlb(\frac{\Db\C}
{z_{10}}-\frac{1}{2}\frac{\theta_{10}}{z_{10}}\D\Db\C).$$

To compute the first integral here we use the following identity
$$\dlb\frac{\Db\C}{z_{10}} = \dlb\Db(\frac{\C}{z_{10}})=
d_{+}(\frac{\C}{z_{10}}) - \dl\D(\frac{\C}{z_{10}})$$ inside ${\cal C}_{k}$.
The term $d_{+}(\frac{\C}{z_{10}})$ does not contribute to $I$ since it is
the total derivative of a single-valued function $(\frac{\C}{z_{10}})$ and the
integration is performed along infinitesimal circles ${\cal C}_{k}$. Whereas
the second term yields
$$-\frac{1}{4}\sum_k\alpha_{0k}\frac{1}{2i\pi}\oint_{{\cal C}_{k}}\dl
(\frac{\D\C}{z_{10}}-\frac{\theta_{10}}{z_{10}^{2}}\C)$$ and this is zero
by the super Cauchy's theorems (\ref{y1}).\\
Thus the integral $J_{2}$ reduces to
$$J_{2} =-\frac{1}{16i\pi}\sum_k\alpha_{0k}\oint_{{\cal C}_{k}}\dlb
\{\frac{\theta_{10}}{z_{10}}[\D\Db\C -
\frac{1}{2}\covder(\D\Hb\D\C) + \D(\zeta_{\theta}(\C\D\Hb + \Hb\D\C))]\}$$
Here the integrand comes with $\theta_{10}$ as a global factor and hence
$J_{2}$ vanishes whatever is the other factor in square brackets due to
the rules (\ref{Br1}) and (\ref{Br3}), or more precisely their complex
conjugates, and the fact that $(\theta_{12})^{2}=0$.

Summing up we have shown that a general Polyakov action on a SRS of
genus $g$ contains three kinds of terms, an integral on $\sig$, a line
integral and a residue contribution
\begin{eqnarray} \label{eb}
\Gamma_{a}[\Hb,\R]&=&\frac{1}{4\pi}\int_{\hat\Sigma} \M
[4(\R-\Rko)\Hb+\D(\Delk+\Delko)\Hb]  \nonumber \\
\Gamma_{b}[\Hb]&=&\fact\oint_{\dsig}\ln(\Psi/\eta_{0})\hat{d}
\ln(\eta/\Psi_{0})\nonumber \\
\Gamma_{c}[\Hb]&=&\frac{1}{4i\pi}\oint_{\dsig}\dlb
\{ (\zeta\zeta_{0}-\chi\chi_{0}-2\zeta\chi)\Hb-
\demi(\chi+\chi_{0}-\zeta-\zeta_{0})\D\Hb\} \nonumber\\
\Gamma_{d}[\Hb]&=&\demi\sum_k\alpha_{0k}\ln(\eta/\Psi_{0})
(P_{0k}).
\end{eqnarray}

This solves the superconformal Ward identity (\ref{j}) on a SRS of higher
genus. The genus which characterizes the SRS appears explicitly in
$\Gamma_{d}$, since $\sum_{k=1}^{N}\frac{\alpha_{0k}}{2}=g-1$, where $N$ is
the number (without counting multiplicity) of zeros of $\eta_{0}$.

Finally we wish to emphasize that by construction this solution is not unique
due to the presence of zero modes, i.e. it is only defined up to addition
of an arbitrary functional ${\cal F}(\Hb)$ satisfying the
condition~~~~$s{\cal F}=0$.

\section{A morphism between the diffeomorphism symmetry and (super)
holomorphy properties}

We note that one can go from eqs.(\ref{k1}) to eqs.(\ref{tl})
by substituting $s$ and $\C$ for $\Db$ and $\Hb$ respectively.
This is a trivial consequence of the fact that the projective
coordinates ($\hat Z,\hat\Theta$) obey the following holomorphic
properties \cite{DG}
\begin{eqnarray}\label{shpc}
\Db\Zh &=& \hat{\Theta}\Db\hat{\Theta}+\Hb(\D\hat{\Theta})^{2}\nonumber \\
\Db\hat{\Theta}&=&-\frac{1}{2}\D\Hb\D\hat{\Theta} +\Hb\partial\hat{\Theta}
\end{eqnarray}
whereas their BRST transformations read
\begin{eqnarray}\label{brspc}
s\hat{Z}&=& -\hat{\Theta}s\hat{\Theta}+\C(\D\hat{\Theta})^{2}\nonumber \\
s\hat{\Theta}&=& \frac{1}{2}\D\C\D\hat{\Theta}+\C\partial\hat{\Theta}.
\end{eqnarray}
Thus substituting in eqs(\ref{shpc}) $s$ for $\Db$ and replacing $\Hb$ by
the superghost $\C$ , we recover eqs.(\ref{brspc}) up to a sign
\footnote{This sign difference follows from the fact that the operator
$\Db$ acts from the left, whereas $s$ acts from the right.}.
Accordingly every function of the superprojective coordinates
will exhibit this relation between its $s$ transformation and its holomorphy
equation. As an example of such objects consider the super affine connection
$\chi$ defined in eq.(\ref{a4}); a particular sample of which is given in
eq.(\ref{B}) (on the torus) where an explicit parametrization in terms of the
superprojective variable $\hat{\Theta}$ is given. The transformation law
(\ref{b1}) and the
holomorphic condition deduced from the first equation (\ref{k1})

\begin{equation}\label{xx1}
\Db\chi=\demi\D\Delk \Hb,
\end{equation}

exhibit the correspondence mentioned above. Other examples of such objects
are the super Schwarzian derivative \cite{FD} and the super Bol operators
\cite{Gie} which are the covariant versions of the  superderivatives
 $\der^{n}\D$ on compact SRS.\\
Since this correspondence involves uniquely the classical fields of the BRST
algebra and not the ghost sector, it is more accurate to speak about a
morphism concerning the gauge symmetry underlying the BRST symmetry.
For instance the nilpotency of the BRST algebra has no analog for the
operator $\Db$. Of course there is an equivalent morphism in the
antiholomorphic sector.\\
Obviously this property remains true in components and in the
bosonic case as well, since the BRST law of the projective coordinate $Z$
\cite{DG}

\begin{equation}\label{x2}
sZ=c\partial Z,
\end{equation}

can be deduced from the Beltrami equation

\begin{equation}\label{xX2}
\bar{\partial}Z=\mu\partial Z
\end{equation}

by replacing the Beltrami coefficient and the partial differential
$\bar{\partial}$ by the ghost $c$ and the BRST operator $s$ respectively.

Finally we mention a recent formulation of $W-$geometry in the light
cone gauge  \cite{Z4} where the same kind of morphism is present.

\section{Projection onto component fields}

In this section we give the expression in components of the super Polyakov
action (\ref{eb}). This action involves superfields whose power series
expansions in the Grassman variables $\theta$ and $\bar{\theta}$ have been
given previously in refs.\cite{DG,AK}. However the analytic properties of
the $\frac{1}{2}-$superdifferentials $\eta_{0}$ necessitate a particular
discussion. In fact, these superfields admit a $\theta-$expansion of the form
\cite{AK}

\begin{equation}\label{ec}
\eta_{0}=\sqrt{\omega_{0}}+i\theta\lambda_{0},
\end{equation}

where $\omega_{0}$ is the $1-$differential on
the underlying Riemann surface $\Sigma$, and $\lambda_{0}$ its
supersymmetric partner. We recall that these fields are holomorphic with
respect to the reference structure.
The analytic behaviour (\ref{t1}) implies for
$\omega_{0}$ the usual algebraic structure expected near the point $z_{0k}$
from the Riemann-Roch theorem namely

\begin{equation}\label{ed}
\omega_{0}(z_{k})=\beta^{2}(z_{k}-z_{0k})^{\alpha_{0k}},
\end{equation}

where $\beta$ is the restriction to $\Sigma$ of $\beta(P_{k})$ in (\ref{t1}),
and $\sum_{k}\alpha_{0k}=2g-2$.
On the other hand this yields the following singular behaviour of the
field $\lambda_{0}$

\begin{equation}\label{ef}
\lambda_{0}(z_{k})=i\theta_{0k}\demi\alpha_{0k}\beta (z_{k}-z_{0k})^
{\demi\alpha_{0k}-1}.
\end{equation}

 Obviously $\lambda_{0}$ and $\omega_{0}$ have to be linked since from
(\ref{t1}) and (\ref{ec}) they share in common some analytical structure
in the vicinity of $z_{0k}$. In fact the $1-$differential
$\lambda_{0}$ behaves like
$iD_{\theta_{0k}}(\sqrt{\omega_{0}})$ in this neighborhood, a behaviour
which is compatible with the analytical properties of differentials on
Riemann surfaces.
 Assuming the expansion:

\begin{equation}\label{eg}
\zeta_{0}=\zeta^{0}_{0}+\theta\zeta^{1}_{0},
\end{equation}

eqs (\ref{ed}), (\ref{ef}) imply

\begin{equation}\label{eh}
\zeta_{0}^{0}=\demi\alpha_{0k}\theta_{0k}\frac{1}{z_{k}-z_{0k}}  \hspace{1cm} ,
\hspace{1cm} \zeta_{0}^{1}=-\demi\alpha_{0k}\frac{1}{z_{k}-z_{0k}}.
\end{equation}

Consequently, since $\zeta_{0}^{1}=-\demi\der\ln\omega_{0}$,
the quantity $\theta_{0k}$ can be interpreted as the ratio of
the residues of the super component of the superaffine connection
and the bosonic  affine connection at the singular point $z_{0k}$.\\
The holomorphic superfield $\R$ admits a $\theta-$expansion of the form

\begin{equation}\label{ei}
\R=\frac{i}{2}\rho_{z\theta}+\theta\demi\cal{R},
\end{equation}

where the bosonic projective connection $\cal{R}$ and its supersymmetric
partner $\rho_{z\theta}$ depend only on the holomorphic variable $z$.
 Moreover, we have

\begin{equation}\label{ej}
\Hb=\bar{\theta}\mu_{\bar{z}}^{z}+\theta\bar{\theta}[-i\alpha_{\bar{z}}
^{\theta}],
\end{equation}

where the spacetime fields $\mu$ and $\alpha$ are the Beltrami coefficient
and its fermionic partner respectively. The $\theta-$expansions of $\eta$,
$\Psi$ and $\Psi_{0}$ are analogous to the expansion (\ref{ec}) with
 coefficients denoted by ($\sqrt{\omega}$; $i\lambda$), ($\sqrt{\Omega}$;
 $i\xi$) and ($\sqrt{\Omega_{0}}$, $i\xi_{0}$) respectively.\\
By substituting the above component field expansions in the first expression
of (\ref{eb}) we find (in the following, we shall simplify the notation by
suppressing all indices on the component fields)

\begin{equation}\label{ek}
\Gamma_{a}(\mu,\alpha;\rho,{\cal R})=\Gamma_{a1}(\mu;{\cal R})+
\Gamma_{a2}(\mu,\alpha;\rho)
\end{equation}

where $\Gamma_{a1}(\mu;{\cal R})$ and $\Gamma_{a2}(\mu,\alpha;\rho)$ are

\begin{equation}\label{el}
\Gamma_{a1}(\mu;{\cal R})=\frac{1}{2\pi}\int_{\Sigma}d^{2}z[2{\cal R}\mu+
2\Dec^{2}\mu
+\partial\ln(\Omega_{0}/\Omega)\Dec\mu
-2\mu\Dec (\partial\ln\Omega)]
\end{equation}

\begin{eqnarray}\label{em}
\Gamma_{a2}(\mu,\alpha;\rho)&=&\frac{1}{2\pi}\int_{\Sigma}d^{2}z[2\rho\alpha+
\frac{\xi}
{\sqrt{\Omega}}\Dec_{0}\alpha+\frac{\xi_{0}}{\sqrt{\Omega_{0}}}\Dec\alpha
-\partial(\frac{\xi}{\sqrt{\Omega}})(\alpha+2\mu\frac{\xi_{0}}
{\sqrt{\Omega_{0}}}) \nonumber \\
&-&\partial(\frac{\xi_{0}}{\sqrt{\Omega_{0}}})(\alpha+2\mu\frac{\xi}
{\sqrt{\Omega}})].
\end{eqnarray}

The functional $\Gamma_{a2}(\mu,\alpha;\rho)$ represents the contributions
which are due to supersymmetry.
In these formulae appears the covariant derivative $\Dec$ associated to the
affine connection $\partial\ln\Omega$ and which is defined by \cite{BFIZ}

\begin{equation}\label{en}
\Dec \equiv\partial -p\partial\ln\Omega
\end{equation}

where $p$ is the conformal weight (relative to the $z-$index) of the field
on which $\Dec$ is applied. When acting on the associated affine connection
or another affine
connection $\partial\ln\Omega '$, the covariant derivative defines the
``field strength''

\begin{equation}\label{eo}
\Dec (\partial\ln\Omega ')\equiv(\partial-\demi p\partial\ln\Omega)
\partial\ln\Omega '.
\end{equation}

The three other contributions of eq. (\ref{eb}) become in components

\begin{eqnarray} \label{ep}
\Gamma_{b}(\mu,\alpha)&=&\fact\oint_{\partial{\cal D}(\omega_{0})}
[\frac{1}{4}\ln(\Omega/\omega_{0})d\ln(\omega/\Omega_{0})
-dz(\frac{\xi}{\sqrt{\Omega}}-\frac{\lambda_{0}}{\sqrt{\omega_{0}}})
(\frac{\xi_{0}}{\sqrt{\Omega_{0}}}-\frac{\lambda}{\sqrt{\omega}})]\nonumber \\
\Gamma_{c}(\mu,\alpha)&=&\frac{1}{4i\pi}\oint_{\partial{\cal D}(\omega_{0})}
d\bar{z}\{(2\frac{\lambda}{\sqrt{\omega}}\frac{\xi}{\sqrt{\Omega}}
-\frac{\lambda_{0}}{\sqrt{\omega_{0}}}\frac{\lambda}{\sqrt{\omega}}
-\frac{\xi_{0}}{\sqrt{\Omega_{0}}}\frac{\xi}{\sqrt{\Omega}})\mu\ \nonumber \\
&+&\frac{i}{2}(\frac{\lambda_{0}}{\sqrt{\omega_{0}}}+\frac{\lambda}{\sqrt{
\omega}}-\frac{\xi_{0}}{\sqrt{\Omega_{0}}}-\frac{\xi}{\sqrt{\Omega}})\alpha\}
\nonumber\\
\Gamma_{d}(\mu,\alpha)&=&\demi\sum_k\alpha_{0k}[\demi\ln(\omega/\Omega_{0})
+i\theta_{0k}(\frac{\lambda}{\sqrt{\omega}}-\frac{\xi_{0}}
{\sqrt{\Omega_{0}}})](z_{0k}).
\end{eqnarray}

In the expressions above are included the results of the bosonic theory
\cite{Z1}. They are obtained by setting $\lambda=\lambda_{0}=\rho=\alpha
=\xi_{0}=\xi=0$. The dissection ${\partial{\cal D}(\omega_{0})}$ was
introduced in \cite{Z1}. The points used to define it are the
zeros of $\omega_{0}$  and since this 1-differential is holomorphic
in the reference conformal structure $\mu=\alpha=0$, this dissection is
independent of $\mu$ and $\alpha$; we recall that $\omega$, $\lambda$,
$\omega_{0}$ and $\lambda_{0}$ are single-valued and have zeros on the
underlying Riemann surface $\Sigma$. On the other hand $\Omega,~\Omega_{0},
{}~\xi,~\xi_{0}$ are multivalued on $\Sigma$ and have no zeros. Explicit
examples of such objects have been given in \cite{Z2} in terms of the
theta function and the prime form.

\section{Concluding comments}

In summary we have derived a general expression for the Polyakov action on
an $N=1$ SRS of arbitrary genus in the resticted geometry $H_{\theta}^{z}=0$
(and c.c). The superfield formulation was obtained through a new formalism
of integration rules reproducing the results of the component expansion while
avoiding its technical complications, thus allowing a complete superfield
treatement throughout the calculations. The main ingredient was the
introduction of a supercontour ( or as we called it ``Grassmann circle'' )
over which the integration with respect to the Grassmann variables
incorporates the usual Berezin rules and also implies the analytic structure
of superfields on the underlying Riemann surface. In addition some technical
difficulties inherent to the supersymmetric approach were circumvented by
using the systematic method of covariant derivatives.\\
As mentioned at the end of paragraph 2, the action we have constructed is only
defined up to a BRST invariant functional. In fact we can go further. Starting
with the action $\Gamma_{a}$ in (\ref{eb}) we replace the multivalued fields
$\chi$ and $\chi_{0}$ by the single-valued ones $\zeta$ and $\zeta_{0}$
respectively. By doing this we end up with an expression which can be shown
to be the first
term for a second solution to the superconformal Ward identity (\ref{j}).
More precisely, let us denote the resulting expression by $\hat A_{\alpha}$
and then by using the holomorphy equations (\ref{k1}) we get
\begin{eqnarray}\label{conc1}
\hat{A}_{1}-\hat{A}_{\alpha}&=&-4\{\Db[\ln(\Psi/\eta_{0})
\D\ln(\eta/\Psi_{0})]
+ \D[\ln(\Psi/\eta_{0})\Db\ln(\eta/\Psi_{0})]\} \\ \nonumber
&-&\D\{ (\zeta\zeta_{0}-\chi\chi_{0}-2\zeta\chi)\Hb-
\demi(\chi+\chi_{0}-\zeta-\zeta_{0})\D\Hb\}
\end{eqnarray}
where $\hat{A}_{1}$ is the integrand in (\ref{a3}).
Next we integrate this expression over $\hat\Sigma$ which is now seen as the
disjoint union of the domains $\hat{{\cal D}}(\eta_{0})$ and
$\hat{{\cal D}}(\eta)$ that contain all zeros of $\eta_{0}$ and $\eta$
respectively, and then use the Stokes theorem (\ref{St}) to obtain the
identity
\begin{eqnarray}\label{conc2}
\Gamma_{\alpha} &=& \Gamma_{a} + \fact\oint_{\dsig}\ln(\Psi/\eta_{0})\hat{d}
\ln(\eta/\Psi_{0}) + \fact\oint_{\partial\hat{{\cal D}}(\eta)}
\ln(\Psi/\eta_{0})\hat{d}\ln(\eta/\Psi_{0})\nonumber \\
&+&\frac{1}{4i\pi}\oint_{\partial\hat\Sigma}\dlb
\{ (\zeta\zeta_{0}-\chi\chi_{0}-2\zeta\chi)\Hb-
\demi(\chi+\chi_{0}-\zeta-\zeta_{0})\D\Hb\}.
\end{eqnarray}

Now we note that in the third term above we have integrable singularities
and thus by doing the same calculation that led to the result (\ref{ab})
 we get
$$\fact\oint_{\partial\hat{{\cal D}}(\eta)}
\ln(\Psi/\eta_{0})\hat{d}\ln(\eta/\Psi_{0})=
-\demi\sum_j\alpha_{j}\ln(\Psi/\eta_{0})(P_{j})$$.

Consequently, the sum $\Gamma_{a}+\Gamma_{b}+\Gamma_{c}+\Gamma_{d}$ is equal
to the sum of the following terms

\begin{eqnarray}\label{conc3}
\Gamma_{\alpha}[\Hb,\R]&=&\frac{1}{4\pi}\int_{\hat\Sigma} \M
[4(\R-R_{\zeta_{0}})\Hb+\D(\Delta_{\zeta}+\Delta_{\zeta_{0}})\Hb] \nonumber \\
\Gamma_{\beta}[\Hb]&=&\demi\sum_k\alpha_{0k}\ln(\eta/\Psi_{0})(P_{0k})
+\demi\sum_j\alpha_{j}\ln(\Psi/\eta_{0})(P_{j}) \nonumber \\
\Gamma_{\gamma}[\Hb]&=&-\frac{1}{4i\pi}\oint_{\hat{{\cal D}}(\eta)}\dlb
\{ (\zeta\zeta_{0}-\chi\chi_{0}-2\zeta\chi)\Hb-
\demi(\chi+\chi_{0}-\zeta-\zeta_{0})\D\Hb\}
\end{eqnarray}
This solution can be seen as the supersymmetric generalization of the second
solution found by Zucchini in \cite{Z2} on a Riemann surface.
One of the advantages of this solution over the one in (\ref{eb}) is the fact
that it can be easily related to the Polyakov action (\ref{h}) we constructed
on the supertorus \cite{AK}. Indeed, a simple calculation yields the following
\begin{eqnarray}\label{conc4}
\hat A_{1} &=& A_{ST} - \D\{\D\Hb\D\ln(\eta/\eta_{0}) -
2\Hb\D\ln\eta\D\ln\eta_{0}\} + 4\D\Db\ln\eta \nonumber \\
&\equiv& A_{ST} + I_{1} + I_{2}
\end{eqnarray}
which holds on a SRS. Then using the fact that $\eta_{0}$ is holomorphic in
the reference structure i.e., $\Db\eta_{0}=0$, we can rewrite $I_{2}$ as

$$I_{2}= 4 \D\Db\ln(\eta/\eta_{0}) $$
thus yielding a well-defined expression since $\eta/\eta_{0}$ is now an
( invariant ) function. $I_2$ is therefore a total derivative of a
single-valued, well-defined and singularity free $(\demi,0)-$superdifferential
$\D\ln(\eta/\eta_{0})$, and hence vanishes upon integrating on small
circles on the supertorus.\\
Similarly, $I_{1}$ is a total derivative of a single-valued and non-singular
$(0,\frac{1}{2})-$superdifferential, since the expression in brackets in
(\ref{conc4}) transforms with the factor $\overline{\D\tilde\theta}$ under
conformal change of coordinates. Thus the integral of $I_1$ over the
supertorus vanishes.
Therefore the restriction of $\hat A_1$ onto the supertorus gives exactly
$A_{ST}$. \\
Now as the differentials on the supertorus have according to the super
Riemann-Roch theorem no zeros, $\Gamma_{\beta}$ vanishes trivially since
$\alpha_{0k}=\alpha_{j}=0$. As to $\Gamma_{\gamma}$ the reasoning is as
follows. Due to the fact that there is a unique holomorphic superdifferential
(in a given structure)
on the supertorus, $\Psi$ and $\Psi_{0}$ become proportional to $\eta$ and
$\eta_{0}$ respectively. The corresponding factors are multivalued
functions, which must be holomorphic on the whole torus
since we want this restriction to hold everywhere thereon; this
implies that they are constant. In this
case $\chi$ and $\chi_{0}$ reduce exactly to $\zeta$ and $\zeta_{0}$
respectively,
and thereby $\Gamma_{\gamma}$ vanishes indentically on the supertorus.\\
There are many issues that deserve serious study, namely the modular
invariance of these solutions and their pertinence to resum the perturbative
series provided by the renormalized field theory as an iterative solution to
the superconformal Ward identity (\ref{j}). This would provide a
generalization of the work done on the supertorus \cite{HKMK}.\\
\vspace{1cm}

{\large\bf Acknowledgments}\\
\vspace{0.5cm}

It is a pleasure to thank A.Sebbar for numerous discussions on Riemann
surfaces and M. Kachkachi for his interest in this work.
We are particularly indebted to R.Zucchini for useful correspondence
about his papers. H.K. acknowledges enlightening discussions with K.S.Narain,
A.Strathdee and R.Zucchini. Finally the authors are grateful to
J.T.Donohue for a careful reading of the manuscript and F.Gieres for valuable
comments on a preliminary version of this work.

\end{document}